\documentclass{article} 
\usepackage{arxiv}
\usepackage{natbib}
\usepackage{wrapfig}            
\usepackage{hyperref}
\usepackage{url}
\usepackage[inline,shortlabels]{enumitem}
\usepackage{amsmath,amsfonts,amssymb}
\usepackage{algorithm,algorithmic}
\usepackage{mathtools}
\usepackage{booktabs}

\usepackage{graphicx}
\usepackage{rotating}
\usepackage{caption}
\usepackage{subcaption}

\graphicspath{{.}}

\usepackage[nameinlink]{cleveref}
\crefname{figure}{Fig.}{Figs.}
\Crefname{figure}{Figure}{Figures}
\crefname{equation}{Eq.}{Eqs.}
\Crefname{equation}{Equation}{Equations}
\crefformat{section}{#2§#1#3}
\crefname{section}{§}{§§}
\Crefname{section}{Section}{Sections}
\crefname{table}{Table}{Tables}
\crefname{appendix}{Appendix}{Appendices}

\title{Explainable Reinforcement Learning-based Home Energy Management Systems using Differentiable Decision Trees}

\author{Gargya Gokhale, \thanks{Under Review} \\
IDLab\\
Ghent university -- imec\\
\texttt{gargya.gokhale@ugent.be} \\
\And
Bert Claessens \\
IDLab, Ghent university -- imec \\
beebop.ai \\
\AND
Chris Develder \\
IDLab \\
Ghent university -- imec \\
}

\begin{document}

\maketitle

\begin{abstract}
With the ongoing energy transition, demand-side flexibility has become an important aspect of the modern power grid for providing grid support and allowing further integration of sustainable energy sources. Besides traditional sources, the residential sector is another major and largely untapped source of flexibility, driven by the increased adoption of solar PV, home batteries, and EVs. However, unlocking this residential flexibility is challenging as it requires a control framework that can effectively manage household energy consumption, and maintain user comfort while being readily scalable across different, diverse houses. We aim to address this challenging problem and introduce a reinforcement learning-based approach using differentiable decision trees. This approach integrates the scalability of data-driven reinforcement learning with the explainability of (differentiable) decision trees. This leads to a controller that can be easily adapted across different houses and provides a simple control policy that can be explained to end-users, further improving user acceptance. As a proof-of-concept, we analyze our method using a home energy management problem, comparing its performance with commercially available rule-based baseline and standard neural network-based RL controllers. Through this preliminary study, we show that the performance of our proposed method is comparable to standard RL-based controllers, outperforming baseline controllers by $\sim20\%$ in terms of daily cost savings while being straightforward to explain.   
\end{abstract}

\section{Introduction}
\label{sec:intro}
The ongoing shift towards sustainable energy is leading to a significant restructuring of the energy sector: large-scale integration of distributed renewable energy sources, increased electrification, phasing out of fossil fuel-based generation, etc.~\cite{irena}. As a result of these changes, there is a growing need for grid balancing services and demand-side flexibility to ensure the reliable and secure functioning of the grid. Conventionally, large industries and big consumers were the primary sources of such demand-side flexibility. However, another important and as-of-yet untapped source of energy flexibility is the residential sector~\cite{iea-2023}. 

However, realizing a solution that unlocks this (residential) flexibility requires effective controllers that can manage the energy consumption of buildings while operating in a sequential way under uncertain operating conditions. Developing such controllers is an extremely challenging task and has been a major research area~\citep{hems-review-3, hems-review-challenges}. A prominent and established method in this domain is Model Predictive Control (MPC). MPC relies on a mathematical model of the system to anticipate its future behavior and an optimizer that uses this model to obtain optimal control actions~\citealp{mpc-basic}. Several works such as~\citep{mpc-hems-2, real-mpc}, demonstrate the effectiveness of MPCs in both simulation and real-world scenarios. However, most MPCs have been limited to large commercial or institutional buildings, owing to their strong dependence on accurate models of the systems~\citep{mpc-problems-1}. 

Consequently, recent research has shifted towards data-driven RL-based methods for designing controllers~\citep{rl-review}. RL-based controllers rely on data obtained by interacting with the household, circumventing the need for bespoke models as common with MPCs. Works such as~\citep{rl-hems-2, safe-rl-hems-1} show such applications of RL in HEMS, with some works also demonstrating real-world pilot studies~\citep{transfer-rl}. While promising, realizing a commercially viable RL-based HEMS is still difficult. One important reason behind this is the lack of explainability associated with RL and deep RL algorithms~\citep{nagy2023ten, xai-energy-review}. 

Since most HEMS are directly exposed to end-users (``ordinary homeowners’’) who may not be energy experts, the primary requirement for any acceptable control policy is how easily can it be explained to end-users. Most RL-based controllers rely on deep neural networks that are inherently black-box and hence difficult to explain~\citep{xai-energy-review}. Additionally, while works in explainable AI such as~\citep{xai-shap-rl, qunat-xai} explore some post-hoc explanation techniques based on SHAP values or feature importance, these explanations are mainly aimed at experts and cannot be offered to ordinary homeowners. 

We identify this as a major obstacle in realizing practical, data-driven, RL-based HEMS and present our work on learning differentiable decision tree~(DDT)-based RL policies as a possible solution to this problem. The key idea is to replace (deep) neural network-based control policy with a simple decision tree-based policy that is structurally explainable i.e., in the form of rather simple \emph{if-then-else} rules, while being able to learn such trees using data and gradient descent.  
Inspired by works such as~\citealp{ddt-rl}, we demonstrate how differentiable decision trees can be used with standard off-policy RL algorithms such as DDPG ~\citep{ddpg} and how trained actors lead to explainable control policies. Concretely, the main contributions of our work are: 
\begin{enumerate}
    \item We introduce a new `actor’ architecture based on differential decision trees to train standard off-policy actor-critic RL agents. 
    \item We demonstrate the usability of such an agent on a preliminary HEMS problem, comparing its performance against baseline and standard RL controllers. 
\end{enumerate}

Note that, while ~\citet{ddt-rl} have previously introduced a similar method, their approach was restricted to Atari games and other benchmark RL domains. To the best of our knowledge, our work is one of the first applications of differentiable decision tree-based RL agents in the energy domain.

\section{Methodology}
\label{sec:method}
We now present our approach of using a differentiable decision tree as an actor in DDPG to learn explainable and simple policies using gradient descent.
\subsection{Problem Formulation}
\label{subsec:problem}
We examine our proposed DDT-based RL controllers in the context of a home energy management system, where the goal is to efficiently control a home battery~(flexibility asset) to optimize the energy bill of a homeowner. For this, we consider an average Belgian household with a rooftop solar PV installation~(with generated power $P^{\text{pv}}_{t}$), non-flexible electrical load~($P^{\text{con}}_{t}$), and a home battery. We assume that this household is exposed to varying BELPEX day-ahead prices~($\lambda^{\text{con}}_{t}$) and a capacity tariff based on peak power~\citep{capacity-tariff}. This leads to a joint optimization problem, where the HEMS must minimize the daily cost of both the energy consumption~($c^{\text{eng}}_{t}$) and the peak power~($c^{p}_{t}$)~(detailed in~\cref{subsec:problem_math}). To better reflect the present-day scenario, we incorporate solar PV and consumption profiles from a real-world household along with the actual BELPEX day-ahead prices. 

\subsection{Reinforcement Learning}
We model the sequential decision-making problem presented in~\cref{subsec:problem} as a Markov Decision Process (MDP)~\cite{rl-basic}. The states~($\mathbf{x}_{t}$) consist of the current price, battery state-of-charge, non-flexible demand~($P^{\text{con}}_{t}$), and solar PV generation~($P^{\text{pv}}_{t}$). The actions~($u_{t}$) are the charging/discharging signals given to the battery. For improved explainability, we assume a discrete action space of 5 elements~(i.e., $\mathbf{U} = \{ -1, -0.5, 0, 0.5, 1\}$), with the possibility of extending it reserved for future work. The reward function~($\rho$) is defined as the cost incurred for each time step $t$ and modeled based on~\cref{subeq:eng_cost}, \cref{subeq:cap_cost}.  The transition function~($f$) models the dynamics of the household and the battery. 

The goal of an RL agent is to find a policy $\pi: \mathbf{X}\rightarrow \mathbf{U}$ that minimizes the expected $T$-step cost starting from an initial state $\mathbf{x}_{0} \in \mathbf{X}$. For our work, we focus on DDPG, a state-of-the-art off-policy, actor-critic algorithm, where the actor learns a control policy and the critic concurrently estimates the optimal state-action value function~($Q$-function). For more details about this algorithm, we refer to~\citep{ddpg}, with additional modifications discussed in~\cref{subsec:discrete_ddpg}.

\subsection{Differentiable Decision Trees}
\label{subsec:ddt}
Differentiable decision trees or soft decision trees are a variant of ordinary decision trees, introduced in prior works such as~\citep{soft-dt-1, soft-dt-2}. Like ordinary decision trees, DDTs have two types of nodes: 
\begin{enumerate*}[(i)]
    \item Decision Nodes; and
    \item Leaf Nodes
\end{enumerate*}
The decision nodes comprise feature selection weights~($\boldsymbol{\beta}$) for selecting a feature and cut-threshold~($\phi$) for splitting across the selected feature. However, unlike ordinary trees, the decision node in DDTs implements a soft decision using the \emph{sigmoid} function~($\sigma$) as shown in~\cref{subeq:decision_node}. The leaf nodes contain an output distribution vector~($\mathbf{w}$), that is tuned to obtain an output probability distribution~($\mathbf{p}^{L}$), which in our case is the probability distribution over the action space~($\mathbf{U}$). This is modeled using \emph{softmax}, where the probability for each action $u_{m} \in \mathbf{U}$ is calculated using~\cref{subeq:leaf_prob}.

\begin{subequations}
\begin{align}
    p^{\text{left}} = \sigma \ (\boldsymbol{\beta} \mathbf{x} - \phi)&  \     ;\ \ p^{\text{right}} = 1-\sigma \ (\boldsymbol{\beta} \mathbf{x} - \phi) \label{subeq:decision_node} \\
    p^{L}_{m} = \frac{e ^ {-w_{m}}}{\sum_{\kappa=1}^{|\mathbf{U}|} e ^ {-w_{\kappa}}} \ & \ \forall m \in \{1, 2, \ldots, |\mathbf{U}|\} \label{subeq:leaf_prob}
\end{align}
\label{eq:ddt-basic}
\end{subequations}

A DDT of arbitrary depth is then built using such decision and leaf nodes~(e.g., ~\cref{fig:ddt_d2}). Each decision node gives the path probabilities for its edges and each leaf node gives the output probability distribution. The final tree is obtained by appropriately combining these probability values following the tree structure. \cref{subsec:app-ddt-2} details the exact formulation of a DDT of depth 2 and its forward pass. At inference, each decision node is converted from the `soft’ version into a `crisp' decision by using \emph{argmax}, \emph{max} operators, resembling an ordinary decision tree. The trainable parameters~($\boldsymbol{\beta}$, $\phi$ and $\mathbf{w}$) are initialized randomly and learned via gradient descent.

\section{Results}
\label{sec:results}
We validate the performance of our proposed DDT-based RL agents on a battery-based HEMS problem~(discussed in~\cref{sec:method}) and investigate the control performance and explainability of the learned controllers. We present the obtained results in this section. 
\subsection{Performance of DDT-based Agents}
\label{subsec:perf}
\begin{wrapfigure}{r}{0.5\textwidth}
\vspace{-0.7cm}
  \centering
    \captionof{table}{Comparison of DDT-based agents}
    \begin{tabular}{c c}
        \toprule
            Algorithm & Mean Cost \\
        \midrule
            DDPG (Standard) &  3.34 € \\
        \midrule
            DDT (depth 2) & 3.47 € \\
        \midrule
            DDT (depth 3) &  $\mathbf{3.02}$  € \\
        \midrule
            Baseline RBC & 4.70 € \\ 
        \bottomrule
    \end{tabular}
    \label{tab:performance}
\end{wrapfigure}
The performance~(mean over 5 seeded runs) of our proposed approach using DDTs of depth 2 and 3 is presented in~\cref{tab:performance}. We note three key observations:
\begin{enumerate*}[(i)]
    \item DDT agents of depth 3 outperform all other agents including standard DDPG;
    \item both DDT agents outperform the baseline RBC controller;
    \item the performance difference between DDTs of depth 2 and standard DDPG agents is quite small ($\sim 4\%$)
\end{enumerate*}.
This indicates that our proposed approach can learn good control policies and outperform typical, built-in RBC included with commercially available batteries. Note that while the mean performance is satisfactory, the DDT-based agents experience some instability in the training process, with some models converging to an inferior control policy (as shown in~\cref{subsec:add-results-perf}). This needs to be investigated further in future work.

\subsection{Explainability of DDT-based agents}
\begin{figure}
    \centering
    \includegraphics[width=\textwidth]{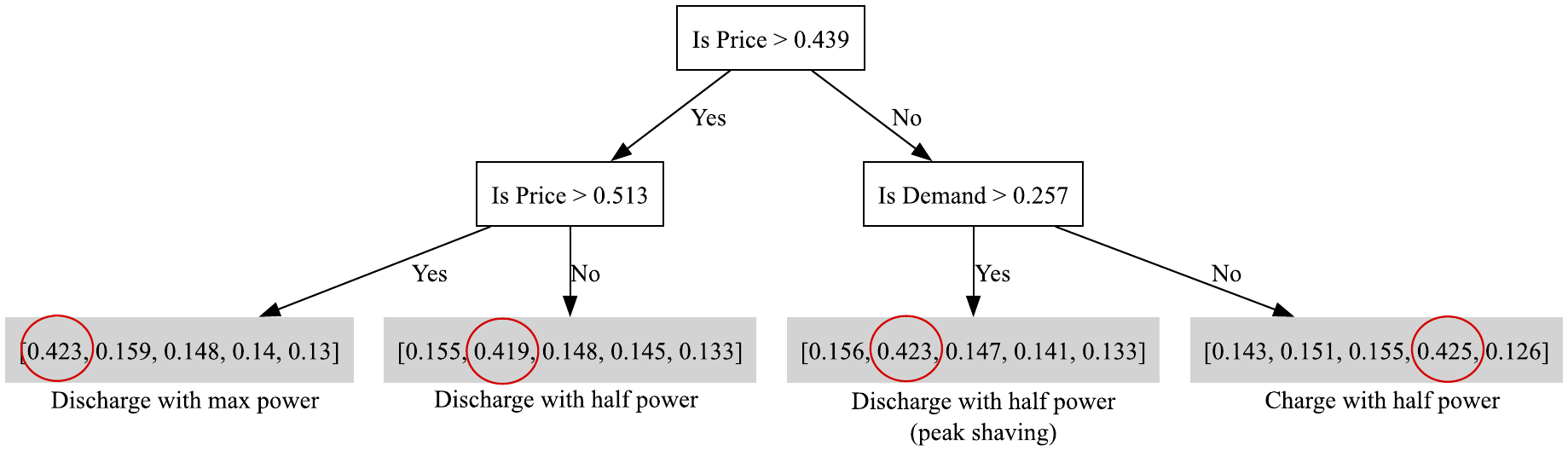}
    \caption{Example of a learned DDT for depth 2 showing selected features, learned thresholds and output distributions. The annotations indicate how the DDT can be explained.}
    \label{fig:ddt_d2}
\end{figure}

In addition to performance, we study the explainability of the learned DDT policies. The learned DDT policies can be easily visualized owing to their tree structure. As an example, \cref{fig:ddt_d2} presents a learned policy of a DDT of depth 2. Note that these DDTs are randomly initialized and over the course of training learn the feature selection~(e.g., choosing `demand' or `price' as the feature for the decision node) and the respective cut thresholds via gradient descent. From~\cref{fig:ddt_d2}, it can be observed that the learned DDT is straightforward to understand and takes intuitive actions -- e.g., the DDT policy only charges the battery when both the price and demand are low, indicating a consistent peak shaving behavior taking into account current as well as future trends. We also present a few other examples of learned policies in~\cref{subsec:add-results-ddts}. 

These preliminary findings clearly show that in addition to the strong performance of the DDTs, the learned policies are also easy to explain and can potentially improve user acceptance of such HEMS. 
 
\section{Conclusion}
\label{sec:conclusion}

With this work, we introduced a novel method for obtaining explainable RL-based control policies using differentiable decision trees. The DDTs can easily `fit' into standard actor-critic RL algorithms as shown in our implementation using DDPG on a battery-based home energy management scenario. As presented in~\cref{sec:results}, our results clearly demonstrate that our proposed DDT-based agents can learn high-quality control policies while being simple and easy to explain. Furthermore, our preliminary findings also show that the DDT-based agents lead to an overall comparable performance as compared to standard neural network-based agents and outperform them in certain settings. 

The goal of this work was to introduce this novel method for explainable RL-based controllers for energy applications. Building on these preliminary results, we aim to further extend our problem scenario, with future work focusing on applying DDTs to develop an elaborate HEMS that can leverage different flexibility assets such as batteries, EVs, and building thermal mass. This involves investigating the performance of such DDT-based HEMS, stabilizing their training and (potentially) deploying them in real-world households to study user acceptance of such an `AI' driven method. 


\bibliography{ref-list}
\bibliographystyle{ACM-Reference-Format}

\appendix

\section{Appendix}

\subsection{Home Energy Management Problem}
\label{subsec:problem_math}
The home energy management problem described in~\cref{subsec:problem} is modeled as:
\begin{subequations}
\begin{align}
\min_{u_{1}, \ldots u_{T}} & \sum_{t=1}^{T} c^{\text{eng}}_{t} + c^{p}_{t} \label{subeq:obj_fun} \\
\text{s.t.:} \ c^{\text{eng}}_{t} &= \begin{cases}
                                \lambda^{\text{con}}_{t} \ P^{\text{agg}}_{t} \ \Delta t &: P^{\text{agg}}_{t} \geq 0      \\
                                \lambda^{\text{inj}}_{t} \ P^{\text{agg}}_{t} \ \Delta t &: P^{\text{agg}}_{t} < 0\\
                                \end{cases} \quad \forall t \label{subeq:eng_cost} \\
                c^{p}_{t} &= \lambda^{\text{cap}} \ \max (P^{\text{agg}}_{t}, \ P^{agg}_{\text{min}})  \label{subeq:cap_cost} \\
                P^{\text{agg}}_{t} &= P^{\text{con}}_{t} + P^{\text{pv}}_{t} + u_{t} \qquad \qquad \ \forall t \label{subeq:u_phys}\\
                E_{t+1} &= \begin{cases}
                                E_{t} + \eta \ u_{t} \ \Delta t &: u_{t} \geq 0      \\
                                E_{t} + \frac{1}{\eta} \ u_{t} \ \Delta t &: u_{t} < 0\\
                            \end{cases} \quad \forall t \label{subeq:bat_mod} \\
                0 \leq E_{t} &\leq E^{\text{max}}; \
                u^{\text{min}} \leq u_{t} \leq u^{\text{max}}  \quad \forall t.
\end{align}
\label{eq:opt}
\end{subequations}

The battery is modeled using a linear model~(\cref{subeq:bat_mod} with charging/discharging actions $u_{t}$ and current energy level~($E_{t}$). The cost of energy consumed~($c^{\text{eng}}_{t}$) depends on the actual power consumed~($P^{\text{agg}}_{t}$) and the current injection and consumption prices~($\lambda^{\text{inj}}_{t}$ and $\lambda^{\text{con}}_{t}$ respectively). Similarly, the capacity cost~($c^{p}_{t}$) depends on the actual power consumed and the minimum power capacity contracted~(which is set to $4\text{kW}$). Furthermore, we assume $T = 24~\text{hours}$ and a time resolution $\Delta t = 1~\text{hour}$. The battery configuration is listed in~\cref{tab:battery_hp}.
\begin{table}[t]
\centering
\caption{Parameters related to the Battery model used in the Home Energy Management Simulator}
\begin{tabular}{cc}
    \toprule
     Parameter                  & Value     \\
    \midrule
    Max Capacity                & 10 kWh     \\
    \midrule
    Max Power                   & 4 kW       \\
    \midrule
   Efficiency (round trip)      & 0.9       \\
   \midrule
   Action Space                 & $\{ -1, -0.5, 0, 0.5, 1\}$ \\
    \bottomrule
\end{tabular}
\label{tab:battery_hp}
\end{table}

\subsection{Formulation of DDT of depth 2}
\label{subsec:app-ddt-2}
\begin{figure}[t]
    \centering
    \includegraphics[width=0.75 \textwidth]{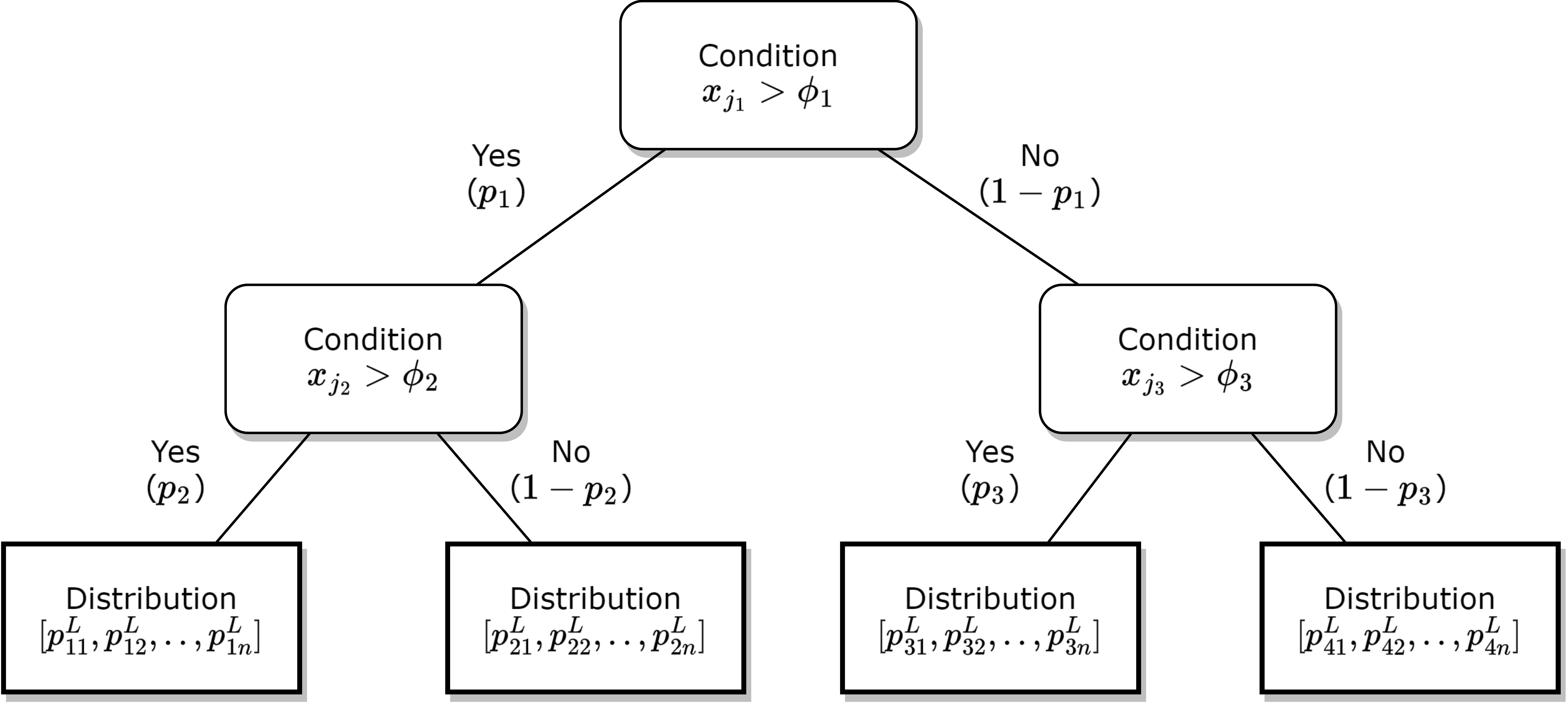}
    \caption{Illustration of a DDT of depth 2 with decision nodes denoted by rounded boxes and leaf nodes with rectangles. Here, all $p_{i}$ represent path probabilities and $p^{L}_{jk}$ represents probabilities at each leaf ($j$) for each element~($k$).}
    \label{fig:formulate_ddt}
\end{figure}

Based on~\cref{subsec:ddt}, we provide the formulation for a DDT of depth 2 (\cref{fig:formulate_ddt}). The path probabilities~($p_{i}$) are computed based on~\cref{subeq:decision_node} and the leaf probabilities~($p^{L}_{jk}$) are computed using~\cref{subeq:leaf_prob}. Algorithm~\cref{alg:ddt_d2} shows the implementation of the DDT. As observed, the path probabilities and leaf probability distributions are first computed. Following this, the probabilities are combined according to the tree structure: 
\begin{enumerate*}[(i)]
    \item For each branch (that starts from the root and ends at a leaf), path probabilities for each edge of the branch and the corresponding leaf node distributions are multiplied to get a probability distribution for that branch;
    \item Output distributions of each of the branches are added to get the final distribution as the output of the DDT.
\end{enumerate*}.
This algorithm is used during the training procedure as the `forward' pass. At inference, all the `soft' operations are converted into `crisp' operations, and the DDT is reduced to an ordinary decision tree. 

\begin{algorithm}[t]
\begin{algorithmic}[1]
\STATE Initialize: $\boldsymbol{{\beta}_{i}}$, $\boldsymbol{\phi}$, $\textbf{w}_{k}^{L}$, where $i = \{1, 2, 3\}$ (decision nodes)  and $k = \{1, 2, 3, 4\}$ (leaf nodes)
\STATE Input: State $\textbf{x}$
\FORALL{i}
    \STATE Feature Selection: $x_{j} = \boldsymbol{{\beta}_{i}} \cdot \textbf{x}  $
    \STATE Evaluate Condition: $p_{i} = \sigma (x_j - {\phi}_{i})$
\ENDFOR
\STATE Calculating Path Probabilities: $ \textbf{p} = 
                            \begin{bsmallmatrix}
                                p_{1} &  0\\
                                0 &  1-p_{1}
                            \end{bsmallmatrix} 
                            \cdot
                            \begin{bsmallmatrix}
                                p_{2} &  1 - p_{2}\\
                                p_{3} &  1-p_{3}
                            \end{bsmallmatrix}
                            $
\FORALL{k}
    \STATE Calculate Leaf Probabilities: $\mathbf{p}^{L}_{k} = \{ p^{L}_{k1}, p^{L}_{k2}, \ldots p^{L}_{kn} \}$ based on~\cref{eq:ddt-basic}
\ENDFOR
\STATE Output: $o = \textbf{p}[1, 1] \textbf{p}^{L}_{1} + \textbf{p}[1, 2] \textbf{p}^{L}_{2} + \textbf{p}[2, 1] \textbf{p}^{L}_{3} +\textbf{p}[2, 2] \textbf{p}^{L}_{4} $
\end{algorithmic}
\caption{Depth 2 DDT Formulation}
\label{alg:ddt_d2}
\end{algorithm}

\subsection{Discrete Actor-based DDPG}
\label{subsec:discrete_ddpg}
As described in~\cref{sec:method}, we use DDTs as the actor-network in DDPG. However, standard DDPG implementations work only with continuous actions. To overcome this challenge, we implement a discrete actor-based DDPG agent. This implementation is inspired by the discrete soft actor-critic implementation presented in~\citep{sac}. The key changes are:
\begin{enumerate*}[(i)]
    \item the target values for critic training are computed using~\cref{eq:critic_target};
    \item the equation for computing actor loss (as gradient of $Q$-values) is modified to~\cref{eq:discrete_actor_loss}
\end{enumerate*}.
In both~\cref{eq:critic_target} and~\cref{eq:discrete_actor_loss}, the computations for value functions are modified, where the sample-based expectation is changed to an explicit calculation of expectation over all actions from action space. 
\begin{equation}
    \text{critic target} = \left( c_{i} + \sum_{k=1}^{|U|} p(u_{k}|\mathbf{x}_{i+1}) \hat{Q}_{\theta^{-}_{c}}(\mathbf{x}_{i+1}, u_{k}) \right)
\label{eq:critic_target}
\end{equation}
\begin{equation}
    \mathcal{L}_{a} = \nabla \ \mathbb{E}\left[  \sum_{k=1}^{|U|} p(u_{k}|\mathbf{x}_{i}) \hat{Q}_{\theta_{c}} (\mathbf{x}_{i}, u_{k})\right]
\label{eq:discrete_actor_loss}
\end{equation}

\subsection{Additional Results: Performance Comparison}
\label{subsec:add-results-perf}
\begin{figure}
    \centering
    \includegraphics[width=0.75\textwidth]{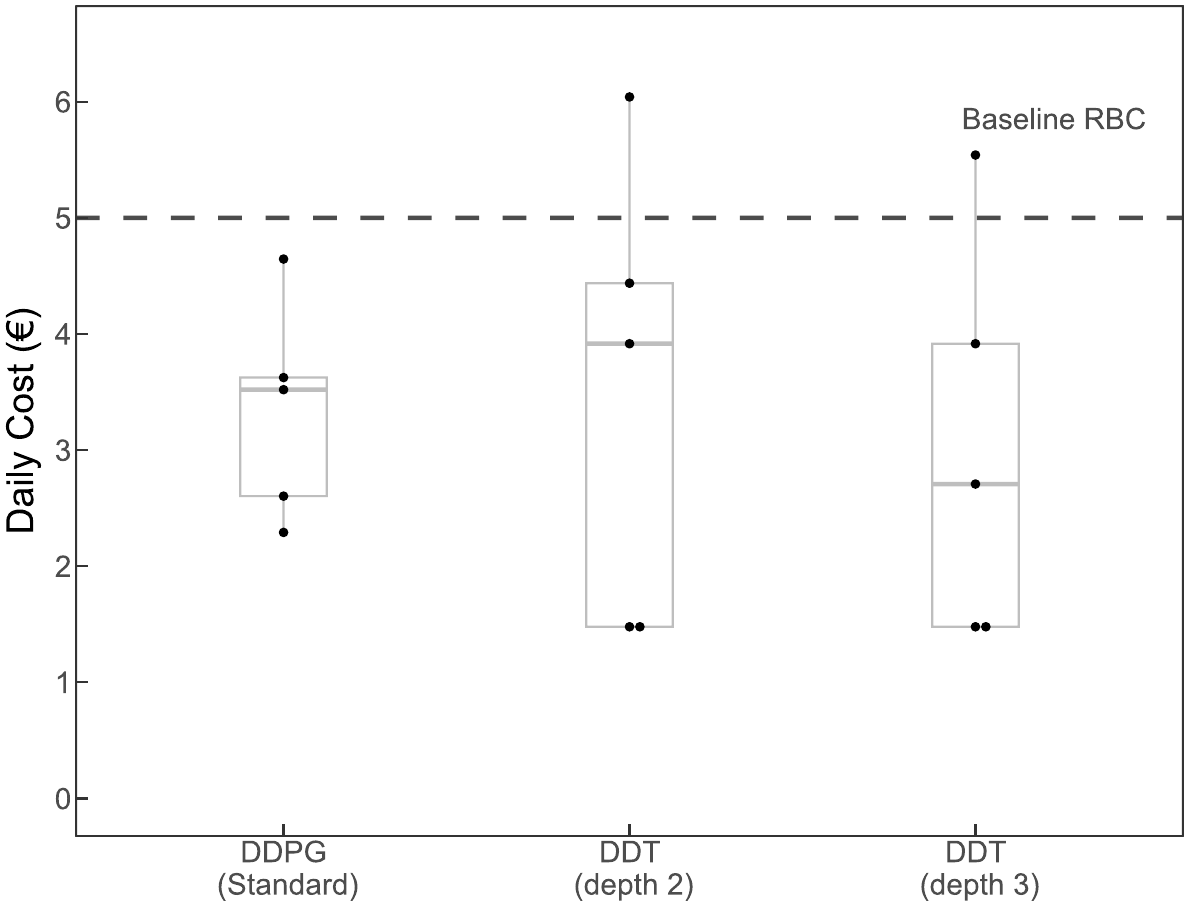}
    \caption{Daily costs attained by DDT-based agents on the HEMS problem. The dots represent the actual performance of individual models and the box plots show the aggregate performance. DDT-based agents are benchmarked using a standard neural network-based DDPG and a baseline RBC.}
    \label{fig:perf_box}
\end{figure}

This section provides additional details related to~\cref{subsec:perf}. 
\label{subsec:add_results}
Section~\cref{subsec:perf} compares the performance of our proposed DDT-based agents with standard DDPG agents and a baseline controller. While~\cref{tab:performance} tabulates aggregate results, ~\cref{fig:perf_box} presents results for all the runs. As discussed in~\cref{sec:results}, some of the DDT-based controllers converge to a sub-optimal value, indicating some instability in the training process. This instability can be attributed to the training process, where changes in `upstream' or hierarchically higher features could have a disproportionate impact on the output distributions. This needs to be investigated further and will be part of future work.

\subsection{Addtional Results: Visualizing Learned DDTs}
\label{subsec:add-results-ddts}
This section provides additional results pertaining to the learned decision trees~(similar to~\cref{fig:ddt_d2}). 
\subsubsection{Depth 2 DDT with redundant rules}
\begin{figure}
    \centering
    \includegraphics[width=\textwidth]{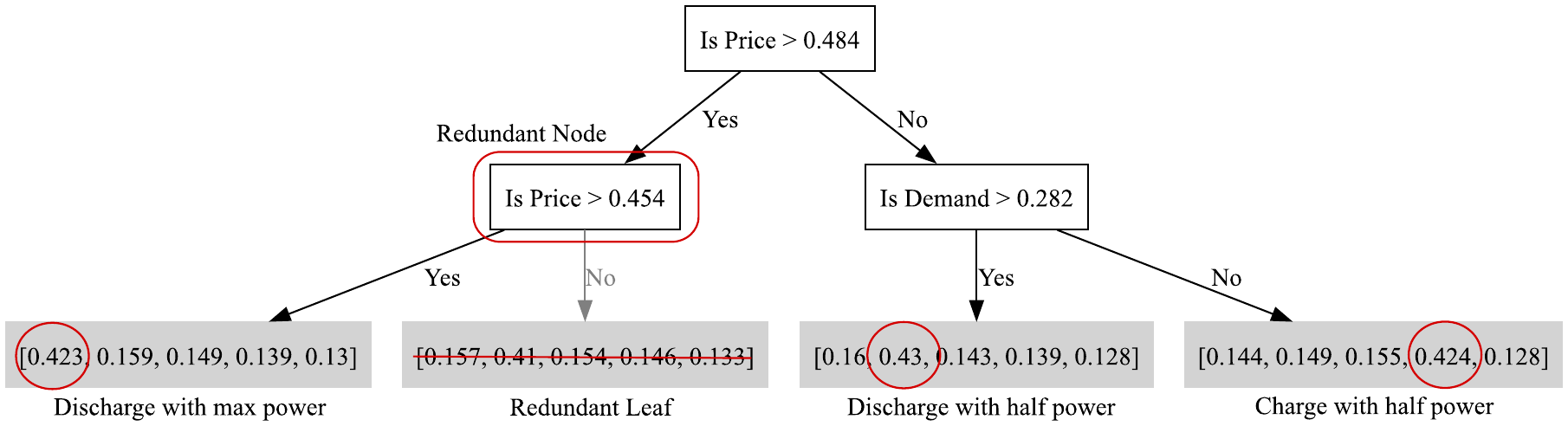}
    \caption{Example of a learned DDT for depth 2 with a redundant decision node}
    \label{fig:d2_redundant_node}
\end{figure}

In some instances, the decision nodes in the DDTs learn redundant or conflicting rules, which in some cases can lead to inferior results. An example of such a DDT is depicted in~\cref{fig:d2_redundant_node}. Here, the highlighted decision node cuts along the same feature as its parent node and learns a redundant threshold leading to one of the leaves being unreachable. This can contribute to training instability and needs to be investigated further. 

\subsubsection{Depth 3 DDT}
\begin{sidewaysfigure}
  \centering
  \includegraphics[width=\textwidth]{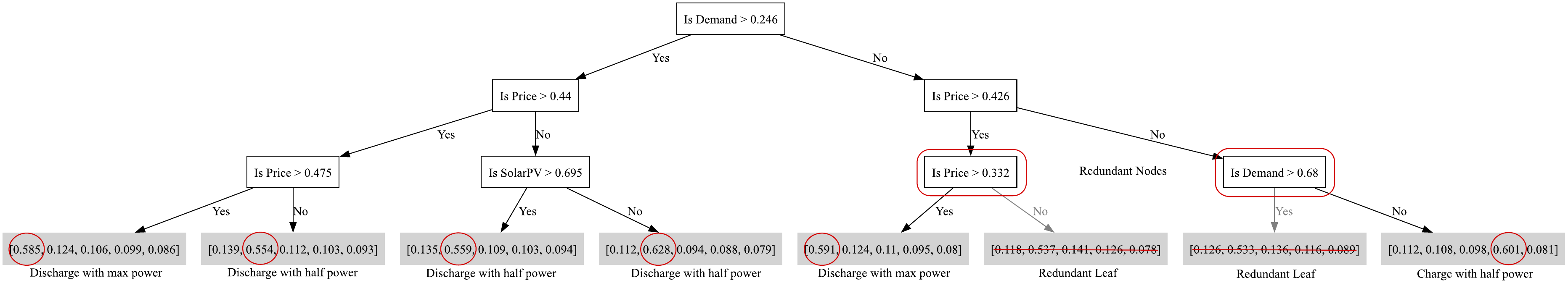}
  \caption{Example of a learned DDT for depth 3}
  \label{fig:ddt_d3}
\end{sidewaysfigure}

Besides DDTs of depth 2, we also trained DDTs with depth 3. An example of such a decision tree is shown in~\cref{fig:ddt_d3}. The higher depth enables this variant to have more decision nodes, leading to a more complex tree representation that can better capture the environment dynamics. However, as observed in~\cref{fig:ddt_d3}, not all leaf nodes are being used due to the decision nodes learning redundant rules, similar to the tree depicted in~\cref{fig:d2_redundant_node}. This indicates the need for further tuning the training process and/or introducing additional loss terms that can penalize such behavior. 

\end{document}